# Influence of DNA on Ag(I)-C-Phycocyanin complexes.

E.Gelagutashvili, E.Ginturi, N.Kuchava, N.Bagdavadze

E.L. Andronikashvili Institute of Physics,
6 Tamarashvili str.,
Tbilisi, 0177, Georgia
E.mail: gel@iphac.ge
gelaguta@yahoo.com

Abstract

The effects of DNA on the fluorescence of Ag(I)- C-phycocyanin (from *Spirulina platensis)* complexes were investigated. It was shown, that the fluorescence intensity of C-phycocyanin decreases, when Ag(I) ions are added but the following DNA addition gives rise to fluorescence buildup.

Fluorescence spectroscopy provides insight of Ag(I) – C-PC interactions. Fluorescence measurements demostrate C-PC quenching of Ag(I) emission intensities. Stern- Volmer quenching constant was obtained from the linear quenching plot. Blue shifts in the fluorescence spectra were observed during Ag(I) binding to C-PC.

## 1. Introduction

Metal ions contribute to the structure and function of many enzymes. The toxic activity of some metals is believed to relate to their ability to compete with essential elements in proteins or to bind to DNA. Silver ions induce errors in DNA transcription processes, that may cause disturbance of normal functionality of nuclear acids. Exhibition of this process may be the bactericidal effect of silver ions as well as mutagenic and cancerogenic activity of $AgNO_3$ [1, 2].

The biocidal effect of silver, with its broad spectrum of activity including bacterial agents, is particularly well known and the term ,, oligodynamic activity" was coined for this phenomenon. Silver ions have an affinity to sulfhydryl groups in enzyme systems, through which they interfere with the transmembranous energy transfer and electron transport of bacterial microorganisms. Silver ions also block the respiratory chain of microorganisms [3]. Blue-green algae Spirulina platensis, like other cyanobacteria, use proteins called phycobiliproteins to harvest light for photosynthesis. Phycobiliproteins, in particular, phycocyanin might be used in photodynamic therapy [4]. Phycocyanin – a natural blue pigment that is the major light –harvesting biliprotein in the blue – green alga Spirulina platensis – performs reduction of normal tissue photosensitivity due to its fast metabolism in vivo [5]. Construction of a marine cyanobacterial strain with increased content of heavy metal ion is tolerant to introducing exogenic metallothionein gene. An expression vector harbouring the C-phycocyanin (C-PC) promoter and C-PC terminal region was constructed and smtA was inserted into its multiple cloning site[6]. The recombinant marine cyanobacteria harbouring the expression vector with smtA could grow even in the presence of 4 μM of $CdCl_2$, the concentration the wild-type strain did not grow in. Marine cyanobacteria, in general, are thought to be particularly sensitive to copper toxicity based on laboratory studies [7]. Phycocyanins extracted from Spirulina are used as industrial and food coloring agents [8]. The blue colour gives the phycocyanins fluorescence properties and the intense fluorescence is exploited in immunoassay tests [9]. Analysis of crystal structure of C-PC packing leads to a proposal for phycobilisome assembly in vivo [10].

In presented work the effect of DNA on the fluorescence of Ag(I)- C-phycocyanin complexes from *Spirulina platensis* using fluorescence spectroscopy is studied.

## 2. Materials and methods

C-PC was isolated from laboratory cultures of the blue -green alga Spirulina platensis, according to Teale and Dale [11]. The extraction of C-PC was carried out in Na –K- phosphate buffer (pH 6,0). Initially the preparation was purified by the method of precipitation in $(NH_4)_2 SO_4$ saturated solution. At the final stage of purification the column, filled with DEAE –cellulose was applied. The buffer solution was at pH 5.2. To determine the degree of purity of the samples spectrophotometry (wave length 250 – 750 nm) and electrophoretic methods were used. The purity of the protein was assessed from the ratio of the absorbances at $\lambda = 615$ nm and $\lambda = 280$ nm ($A_{615} / A_{280}$ >4). The concentration of C –PC was determined by UV/ Vissible spectroscopy using a value of $\varepsilon_{\lambda=615\,nm} = 279\,000\ M^{-1}\ cm^{-1}$ for the absorption coefficient. Other reagents used (AgNO$_3$, NaNO$_3$), DNA („serva") all of analytical grade were prepared in double distilled water. Fluorescence spectra were measured in 1cm$^3$ quartz cells using a fluorescence spectroscopy. Fluorescence titration was performed in the range 400 – 700 nm, by adding a silver solution to C-PC and recording the spectrum after each addition. The solution were excited at 488 nm and the fluorescence intensity was monitored at 635 nm.

### 3. Results and Discussion

In fig.1 are presented fluorescence spectra of 0.40 μM C-PC with increasing concentration of Ag (I) ( 2÷8 μM ) and excitation at 488 nm . The intensity of C-PC, observed in the range 580 – 690 nm, was monitored as a function of added Ag(I) ions.

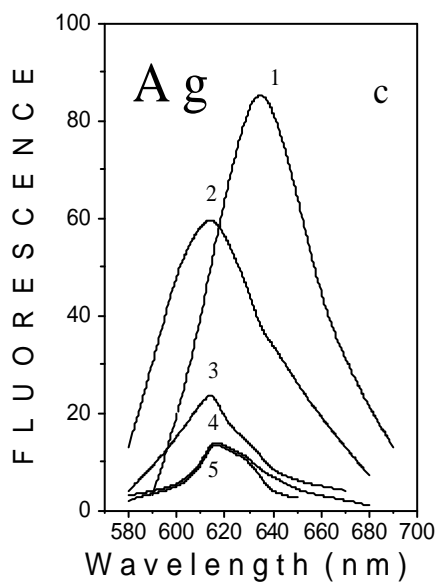

Fig.1. Fluorescence spectra of 0,4 μM C –PC in the presence of different concentrations of Ag(I). (C$_{Ag}$ - 2 ÷8 μM). Wave length of excitation = 488 nm, monitoring at 635 nm.

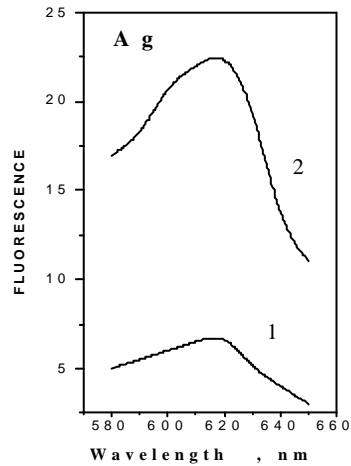

Fig.2 Fluorescence spectra of 0, 4 µM C –PC in the presence of Ag(I). ($C_{Ag}$ - 2 ÷ 8 µM), when amplification is 3,5 (1) and 8,5 (2).

It is seen, that the fluorescence intensity of C-PC decreases with increasing Ag(I) concentration . It is clear that quenching was not saturated even at high concentrations of Ag(I).

In fig.2 are presented fluorescence spectra of 0.40µM C-PC in the presence of Ag(I) ($C_{Ag}$ = 8µ M), when amplification is 3.5 (1) and 8.5 (2). Lack of saturation in the quenching plot even when the binding is saturated indicates that only a part of the binding sites quench the fluorescence.

Normalized fluorescence spectra of C-PC in the absence (1) and in the presence (2) of 5µM Ag(I) (fig.3) shows, that a important blue shift (15 nm) fluorescence is observed in the presence of Ag(I) ions.

Thus, in the presence of Ag(I) ions the fluorescence spectra of C-PC change dramatically showing strong decrease in the peak intensities and blue shifts of the bands. It should be noted, that fluorescence quenching tryptophan and its derivatives by Ag(I) has been observed previously [12] and fluorescence quenching of C-PC by organic compounds, pH, temperature was described in [13].

Stern – Volmer quenching constant ($K_{SV}$) obtained from the linear quenching plot
( fig.4).

$$I_o/I = 1 + K_{SV} [Ag], \quad (1)$$

where $I_o$ and $I$ are the fluorescence intensities in the absence and in the presence of Ag. The data according to eq.1 resulted in a quenching constant of
$K_{SV} = (0.25 \pm 0.02) \times 10^6$ M$^{-1}$ (fig. 4.).

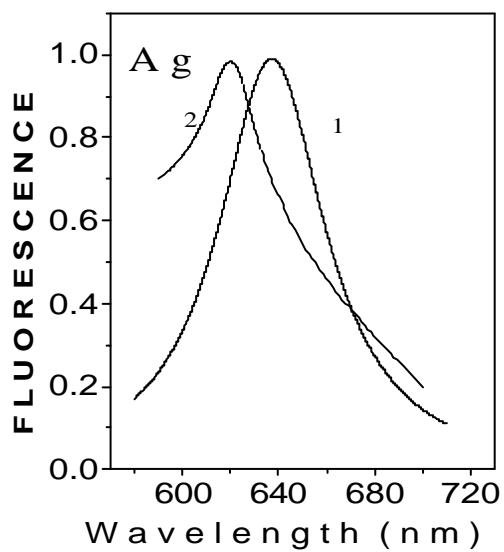

Fig.3. Normalized fluorescence spectra of 0,4 µM C-PC (1) in the absence and (2) in the presence of Ag(I). ($C_{Ag}$ = 7µM).

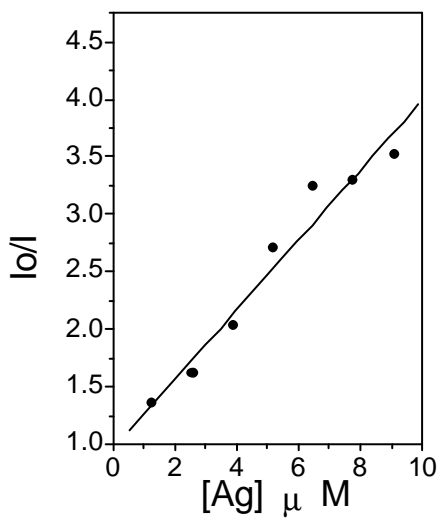

Fig.4. Stern – Volmer plot for quenching of C – PC by Ag(I).

It is known, that the prosthetic group of C-PC is phycocyanobilin (PCB), a linear tetrapyrrole that makes the native protein brigtly coloured and fluorescent [14]. Phycobiliproteins carry linear tetrapyrrole chromophores (bilins) thiother-linked to specific cysteine residues. (C-PC: $(\alpha\beta)_6 - 18$ PCB, where the amino acid sequence is the same in the $\alpha$ and $\beta$ subunits of the phycobiliproteins [15]). It is clear from the present study that every PCB binds one Ag i.e. these data form on the silver-specific nature of thiol groups in C-PC from S.platensis. It is known, that metallothionein is a small protein that binds metal cations by means of numerous cysteine thiolates and the first bacterial metallothionein has been characterized in cyanobacteria [16]. The metal-specific nature for Ni, Cu and Hg of thiol and exopolymer production in a cyanobacterium (Nostoc spongiaeforme) was discussed in [17]. In [18] was shown that the covalently linked chromophore, phycocyanobilin, is involved in the antioxidant and radical scavenging activity of phycocyanin. In [19] the effects of Cr, Pb, Ni and Ag on growth, pigments, protein and al. of *Nostoc muscorum* is studied. The statistical tests revealed a direct positive correlation between the metal concentration and inhibition of different processes. Thiols groups are the target for a number of metal - enzyme interactions [20, 21]. Investigations on Na /K -ATPase demostrate that silver ions inhibit enzyme activity. This inhibition is reversible with the introduction of cysteine [3].

The effect of DNA on the interaction of C-PC – Ag(I) has been investigated also by fluorescence spectrometry method . Fluorescence spectra of C-PC and C-PC + Ag(I) are given in Fig. 5 As it can be seen, the fluorescence intensity of C-PC decreases when Ag(I) ions are added but the subsequent DNA addition gives rise to fluorescence buildup. After repeated DNA addition extra buildup is not observed. As it can be seen from the figure, DNA causes the red shift of fluorescence intensity maximum compared to that of C-PC – Ag complex and location of fluorescence intensity maximum of C-PC – Ag – DNA is the same as that of C-PC. Proceeding from this fact, we may suppose that either DNA binds part of Ag(I) ions, [the remainder of Ag(I) ions is still bound to C-PC, as the fluorescence intensity maximum of complex (3) is less than that of pure C-PC i.e. part of fluorophores remains quenched at the expense of binding of C-PC – Ag] or the other type of interaction is possible: Ag(I) ions simultaneously bind to DNA and C-PC [an intermolecular complex of chelate type C-PC – Ag – DNA is formed], that results in disturbance of thermodynamic balance of C-PC – Ag complex.

The fluorescence spectra of of 0.4 $\mu$M C-PC +7$\mu$M Ag(I) immediately after complex formation (1) and a day later (2) are shown in fig.6. As it can be seen, the decrease of fluorescence intensity is a function of time. It enables us to assume that such conformational changes of protein occur that all the fluorophores become approachable for Ag(I).

On the basis of presented data we can conclude that although Ag(I) interaction with C-PC is stronger, DNA affects Ag(I)-C-PC complex.

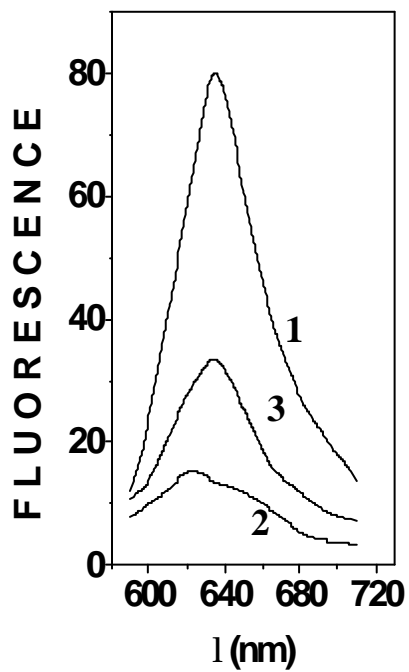

**Fig. 5.** Fluorescence spectra of
 3.2 µM C-PC (1);
 3.2 µM C-PC + 5 µM Ag(I) (2);
 3.2 µM C-PC + 5 µM Ag(I) +
 + 100 µM DNA (3).

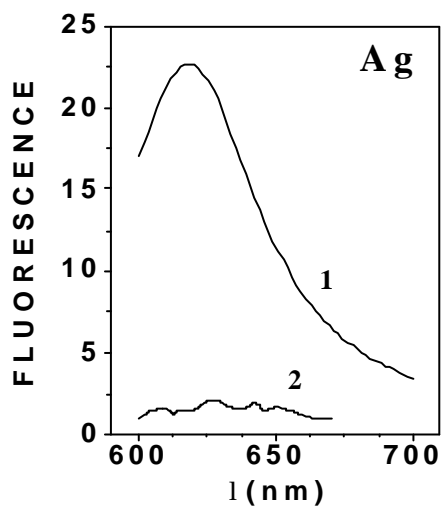

Fig.6. Fluorescence spectrum of 0.4µM C-PC +7µM Ag(I) (1);
 And the same spectrum a day later (2);